\documentclass[11pt,aps,nofootinbib,amssymb,preprint]{revtex4-1}

\usepackage{amsmath,setspace,subfigure,amsfonts,latexsym}
\usepackage{amssymb}
\usepackage{color}
\usepackage{epsfig}
\usepackage{graphics}
\usepackage{sidecap}
\definecolor{White}{rgb}{1,1,1}
\definecolor{Red}{rgb}{1,0.1,0}
\definecolor{LightYellow}{rgb}{1,1,.875}
\definecolor{SteelBlue}{rgb}{.273,.508,.703}
\definecolor{navy}{rgb}{0,0,.5}
\definecolor{LightCyan}{rgb}{.875,1,1}
\definecolor{DarkRed}{rgb}{.543,0,0}
\definecolor{HotPink}{rgb}{1,.41,.70}
\definecolor{ForestGreen}{rgb}{.13,.54,.13}
\definecolor{OliveDrab}{rgb}{.42,.55,.14}
\definecolor{MediumBlue}{rgb}{0,0,.80}
\definecolor{RoyalBlue}{rgb}{.25,.41,.88}
\definecolor{DeepSkyBlue}{rgb}{0,.746,1}
\definecolor{Brown}{rgb}{0.545,0.271,0.074}

\def\bea{\begin{eqnarray}}
\def\eea{\end{eqnarray}}
\def\bec{\begin{center}}
\def\ec{\end{center}}

\def\beq{\begin{equation}}
\def\eeq{\end{equation}}

\newcommand\lsim{\mathrel{\rlap{\lower4pt\hbox{\hskip1pt$\sim$}}
    \raise1pt\hbox{$<$}}}
\newcommand\gsim{\mathrel{\rlap{\lower4pt\hbox{\hskip1pt$\sim$}}
    \raise1pt\hbox{$>$}}}
\def\bea{\begin{eqnarray}}
\def\eea{\end{eqnarray}}
\def\ba{\begin{array}}
\def\ea{\end{array}}
\def\bc{\begin{center}}
\def\ec{\end{center}}
\def\nn{\nonumber}

\def\la{\langle}
\def\ra{\rangle}

\begin{document}

\title{\Large 125 GeV Higgs as a pseudo-Goldstone boson \\
in supersymmetry with vector-like matters}

\author{
Kyu Jung Bae$^1$
\footnote{kyujung.bae@kaist.ac.kr},
Tae Hyun Jung$^2$ 
\footnote{thjung0720@gmail.com}
and Hyung Do Kim$^2$
\footnote{hdkim@phya.snu.ac.kr}
}
\affiliation{{\it $^1$Department of Physics, KAIST, Daejeon 305-701, Korea\\
$^2$Department of Physics and Astronomy, Seoul National University, Seoul 151-747, Korea}}
\begin{abstract}

We propose a possibility of the 125 GeV Higgs being a pseudo-Goldstone boson in supersymmetry with extra vector-like fermions. Higgs mass is obtained from loops of top quark and vector-like fermions from the global symmetry breaking scale $f$ at around TeV. The $\mu, B\mu/\mu \sim f$ are generated from the dynamics of global symmetry breaking
and the Higgs quartic coupling vanishes at $f$ as $\tan \beta \simeq 1$.
The relation of  $m_{\rm soft} \sim 4\pi M_Z$ with $f \sim \mu \sim m_{\rm soft} \sim$ TeV is obtained
and large $\mu$ does not cause a fine tuning for the electroweak symmetry breaking.
The Higgs to di-photon rate can be enhanced from the loop of uncolored vector-like matters.
The stability problem of Higgs potential with vector-like fermions can be nicely cured by the UV completion with the Goldstone picture. 

\end{abstract}

\maketitle

\section{Introduction}

The ATLAS and CMS group of the LHC experiments announced 5 $\sigma$ discovery of the Higgs boson from the analysis of various channels \cite{:2012gk, :2012gu}.
The invariant mass distribution from di-photon and 4 leptons of $Z Z^*$ showed a narrow peak at around 125 $\sim$ 126 GeV. The enhanced di-photon event observed in 2011 has been reduced in 2012. The number of signal events in the combined di-photon analysis is slightly larger than the Standard Model prediction though it is compatible within 2 $\sigma$ variance. 
In this paper we look for the implication of the measured Higgs mass
and di-photon excess in the Higgs search with no discovery of supersymmetric particles.

The Higgs boson discovery came earlier than expected
while weak scale supersymmetry discovery is delayed.
The usual expectation before the running of the LHC was the opposite
that supersymmetric particles were expected to be discovered before the Higgs boson 
if supersymmetry is relevant for the electroweak symmetry breaking.
The current search limit for the gluino and squark is above TeV
and pushes the supersymmetric models to the corner of the parameter space
\cite{:2012cwa}.

This tension and the puzzle recently motivated the natural supersymmetry 
\cite{Dimopoulos:1995mi, Cohen:1996vb, Pomarol:1995xc, Dermisek:2006ey, Kitano:2006gv, Asano:2010ut, Papucci:2011wy, Hall:2011aa}.
The stop, the supersymmetric partner of the top quark can be hidden in the top background
if the stop mass is at around 200 GeV and the search can be difficult
\cite{Draper:2011aa, ATLAS:2012ah, :2012pq}.
Degenerate spectrum can be an alternative due to much weaker bounds from the search
as the bound in this case would come from the mono-jet search which is $\alpha_s$ suppressed 
\cite{Dermisek:2006qj, Bae:2007pa, LeCompte:2011fh, Belanger:2012mk, Dreiner:2012gx}.
Nevertheless, in general supersymmetry is pushed to high scale
and natural electroweak symmetry breaking from weak scale supersymmetry
looks more and more difficult to achieve.
This is also supported by the absence of chargino and neutralino search
which is predicted to be light if $\mu$ is small.

We propose a setup in which the electroweak symmetry breaking scale
is lower than the supersymmetry breaking scale.
The $\mu$ problem in supersymmetric models
\cite{Kim:1983dt} has a nice solution 
only in gravity mediation \cite{Giudice:1988yz}.
It lies in the center of the electroweak symmetry breaking
and make the model building with other mediation mechanism quite difficult
\cite{Dvali:1996cu, Choi:1996vz, Choi:1999yaa, Giudice:2007ca}.
We consider a model with large $\mu$ without spoiling natural electroweak symmetry breaking.
\footnote{Higgs phenomenology of large $\mu$ with light CP odd Higgs has been studied
in \cite{Carena:1995bx, Kim:2009sy}.}
The Higgs appears as a pseudo-Goldstone boson of the global symmetry breaking at around a few TeV and the electroweak symmetry breaking scale is independent of $\mu$,
more precisely $M_Z^2$ is one loop suppressed compared to $\mu^2$, $M_Z \sim \mu/4\pi$.
At the scale of the global symmetry breaking, the Higgs potential is absent
if the global symmetry was exact at the symmetry breaking scale.
The global symmetry is explicitly broken by top Yukawa coupling and gauge couplings.
As a result there appears a loop correction from the global symmetry breaking scale 
down to the weak scale.

There are several attempts along this direction with the idea of Higgs being a pseudo-Goldstone boson \cite{Kaplan:1983sm, Georgi:1984af, ArkaniHamed:2002qx, ArkaniHamed:2002qy} and also in the supersymmetric context
\cite{Birkedal:2004xi, Chankowski:2004mq, Berezhiani:2005pb, Roy:2005hg, Csaki:2005fc, Falkowski:2006qq, Bellazzini:2008zy, Bellazzini:2009ix, Kaminska:2010pc}.
We emphasize that there are two important aspects which are not discussed in the literature.
Firstly, $\mu$ is generated at $f$, the global symmetry breaking scale.
Secondly, the presence of new Yukawa couplings with extra vector-like fermions can explain the 125 GeV Higgs boson mass even for $\tan \beta =1$ at the scale $f$.

Double protection of Higgs mass with supersymmetry and global symmetry 
generates a small quartic generated by top loop from $f$ to $m_t$ (the top mass)
and predicts a very light Higgs mass which is incompatible with the obseved Higgs mass of 125 GeV.
We show that this problem can be overcome if there are extra vector-like matters
with order one Yukawa couplings with the Higgs.
If the vector-like fermion is at around a few hundreds GeV and sfermion is at TeV,
the extra vector-like states provide extra loop correction comparable to top loop
\cite{Moroi:1991mg, Moroi:1992zk, Maekawa:1995ha, Babu:2008ge, Martin:2009bg, Graham:2009gy, Moroi:2011aa, Martin:2012dg} and can make the Higgs as heavy as 160 GeV.
Note that 
\bea
(160 {\rm GeV})^2 \sim 3 M_Z^2 = M_Z^2 ({\rm tree}) + 2M_Z^2({\rm loop}). \nn
\eea
Therefore, the entire Higgs mass
\bea
m_h^2 = (125 \ {\rm GeV})^2 \sim 2 M_Z^2, \nn 
\eea
can come from the loop correction from $\mu \sim f$ even if the tree level mass vanishes.

The vector-like fermions with order one Yukawa couplings can enhance the Higgs to d-photon rate compared to the Standard Model \cite{Carena:2012xa, Blum:2012ii, Joglekar:2012hb, ArkaniHamed:2012kq, Kearney:2012zi, Ajaib:2012eb}.
However, it suffers from vacuum stability
if the theory is extrapolated to high energy as order one Yukawa couplings make
the Higgs quartic to be negative at high energy.
The scale of transition can be as low as TeV if the Yukawa coupling of the vector-like fermions is of order one.
In this paper we show that the rapid running of the Higgs quartic coupling
driven by Yukawa couplings with the vector-like fermions
is a virtue in the Goldstone boson picture rather than a problem.
With the running, the Goldstone Higgs boson can be as heavy as 125 GeV.

\begin{figure}[t]
\includegraphics[width=0.95\textwidth]{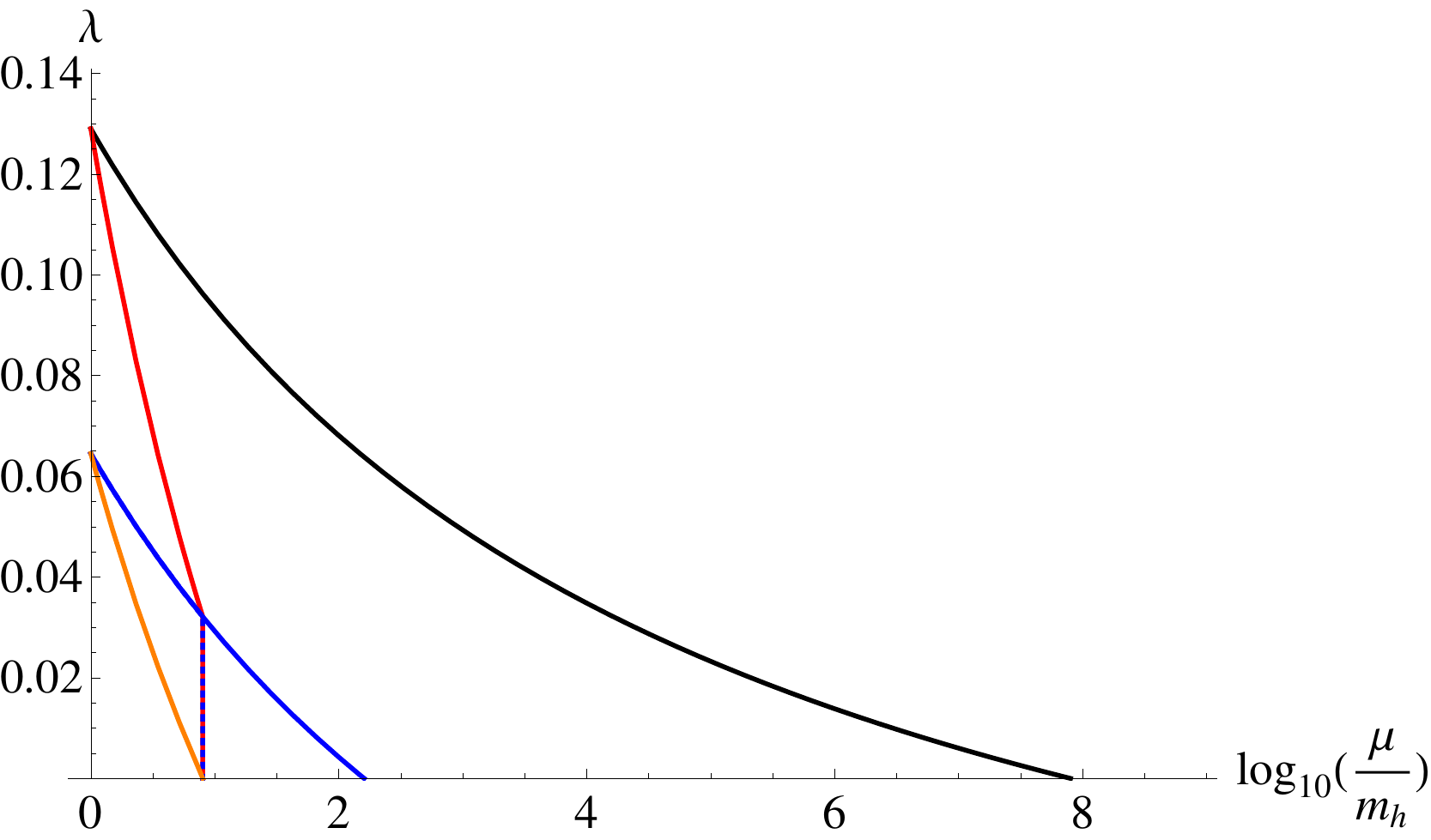}
\caption{Schematic plot for the running of the Higgs quartic coupling $\lambda$.\\
The black line corresponds to the running of the Standard Model Higgs quartic.
The change of the slope is mainly due to the running of the top Yukawa coupling
which becomes small at high energy.
The blue line shows the running of $\lambda$ starting at around 10 TeV down to $m_h$ by top quark Yukawa for $\tan \beta=1$ and no finite threshold correction from the stop.
The Higgs mass is  $m_h^2 \simeq M_Z^2$ in this case. 
In the presence of the stop threshold correction (vertical line at around 1 TeV),
the Higgs mass $m_h^2 \simeq M_Z^2$ is obtained with stop mass 1 TeV for $\tan \beta =1$
(Half from finite threshold correction and the other half from logarithmic running of top quark).
The orange line shows the rapid running of $\lambda$
in the presence of vector-like fermions without top/stop contributions.
Starting from vanishing quartic  ($\lambda=0$) at TeV, $m_h^2 \simeq 2M_Z^2$ is obtained
with the help of top/stop and vector-like matters which is shown in red line.}
\label{Fig-1}
\end{figure}

Goldstone boson picture predicts too light Higgs mass from top loop alone.
Vector-like fermions needed to explain the di-photon enhancement make the Higgs potential unstable at high energy.
These two problems are cured if the Goldstone boson explanation comes with the extra vector-like fermions. Higgs mass can be as large as 125 GeV with the aid of the vector-like fermion Yukawa couplings. The stability problem was due to the simple extrapolation of the low energy theory up to high energy. The Higgs quartic is zero at $f$ and new degrees of freedom appear from the scale $f$.
The simple extrapolation to higher energy is failed. Indeed the potential is flat if the global symmetry is restored. The theory is linked to the supersymmetric theory in which the potential is bounded from below. Goldstone picture and the vector-like fermions are the perfect combination which make the setup phenomenologically viable.

This setup provides an interesting example in which the natural electroweak symmetry breaking
is not directly related to $\mu$ but is smaller by a loop factor. No discovery of light chargino and neutralino would be understood in this setup.

The paper is organized as follows.
First, we explain the basic idea of generating large $\mu$
and also having Higgs as a pseudo-Goldstone boson in supersymmetry.
Second, we introduce the models with vector-like supermultiplets.
Third, we consider a loop correction to the Higgs potential with vector-like matters.
Fourth, we compute the Higgs to di-photon event rate and compare it with the recent LHC data.
Finally, we conclude that the pseudo-Goldstone Higgs is a viable option in supersymmetry
if extra vector-like matters are present and the stability issue turns out to be a merit
to provide the measured Higgs mass.

\section{A pseudo-Goldstone boson in supersymmetry}

In this paper we do not discuss how the global symmetry is realized at a few TeV
from UV theory with higher cutoff and leave it as a separate topic of future work.
Also we do not discuss the vacuum alignment problem.
Here the global symmetry breaking occurs along the direction orthogonal to the electroweak symmetry and the electroweak symmetry breaking appears from the dynamics of the Goldstone boson.
When the global symmetry is spontaneously broken by the vacuum expectation value of the singlet,
a Goldstone boson appears.
If the $\mu$ term is generated in connection with the global symmetry breaking,
we can separate the electroweak symmetry breaking from the $\mu$ parameter
which is tied to the global symmetry breaking scale.
The electroweak symmetry breaking occurs from the loop correction of the pseudo-Goldstone
boson potential.
As a result the electroweak symmetry breaking scale can be naturally low compared to the $M_Z \sim\mu/4\pi$ without causing severe fine tuning of the parameters.

There are various versions of models in which the Higgs boson appears
as a Goldstone boson in supersymmetry.
In most cases the generation of $\mu$ is not discussed in detail in the model construction.
If $\mu$ is very small compared to other soft supersymmetry breaking parameters,
this can be accepted as a good approximation.
However, many model buildings suffer from the generation of the right size $\mu$
and also $\mu/B\mu$ \cite{Dvali:1996cu, Choi:1996vz, Choi:1999yaa, Giudice:2007ca}.
Indeed Giudice-Masiero mechanism in gravity mediation 
is the only natural solution and all of the other mediation models suffer from $\mu/B\mu$ problems.
Therefore, it would be appropriate to discuss the generation of $\mu$ from the beginning.

The mechanism of Goldstone boson Higgs in supersymmetry with $\mu$ generation 
can be general and can be realized in many different concrete models
but in this paper we take a simple specific example to illustrate how it works.
The simple example is provided in \cite{Dvali:1996cu}.

The superpotential is given as follows.
\bea
W & = & S ( \lambda_1 H_u H_d + \lambda_2 N_H \bar{N}_H + \lambda_3 \Phi \bar{\Phi} - M_N^2)
+ \lambda_X X \Phi \bar{\Phi},
\eea
where all of the couplings including $M_N$ are taken to be real by field redefinition.
$H_u$ and $H_d$ are two Higgs doublet superfields, 
$N_H$ and $\bar{N}_H$ are two singlet superfields and $\Phi$ and $\bar{\Phi}$ are the messengers belong to $5$ and $\bar{5}$ representation of $SU(5)$.
$X$ is the spurion for the supersymmetry breaking with $\langle \lambda_X X \rangle = M + \theta^2 F$.
For the special choice of couplings $\lambda = \lambda_1 = \lambda_2$, a global $SU(3)$ symmetry appears.
The fields $\Sigma = (H_u,N_H)$ and $\bar{\Sigma} = (H_d, \bar{N}_H)$ transform as triplet and anti-triplet under the $SU(3)$ transformation. In the $SU(3)$ limit, the superpotential is written as follows.
\bea
W & = & S ( \lambda \Sigma \bar{\Sigma} + \lambda_3 \Phi \bar{\Phi} - M_N^2)
+ \lambda_X X \Phi \bar{\Phi},
\eea
In the supersymmetric limit, the VEVs of $N_H$ and $\bar{N}_H$ break the $SU(3)$ down to $SU(2)$.
Gauge and Yukawa interactions break $SU(3)$ symmetry explicitly.
There are $9-4=5$ broken generators corresponding to one Higgs doublet and one singlet and the corresponding Goldstone bosons get their masses from loops.
In principle, it is possible that the $SU(2)$ doublet $H_u$ and $H_d$ can develop the vacuum expectation value (VEV) just like $N_H$ but we do not consider this vacuum alignment problem in this paper and assume that only the $N_H$ and $\bar{N}_H$ get VEVs.

At one loop, a tadpole for $S$ appears in the potential after integrating out the messengers
and $\la S \ra \neq 0$ is developed.
The tadpole for $S$ arises only after the supersymmetry breaking and is absent
in the supersymmetric limit.
As a result of non-zero vacuum expectation value (VEV) of $S$ ($\la S \ra$),
the F-flat relation for $S$ is not satisfied and $B\mu = \lambda_1 \la F_S \ra$ is generated, too.
\bea
\mu & = & -\frac{5}{32\pi^2} \frac{\lambda_1 \lambda_3 \lambda_X}{\lambda_2} \frac{F^2}{M_N^2 M},  \\
B\mu & = & -\frac{\lambda_2}{\lambda_1} \mu^2,
\eea
where $M=\lambda_X \langle X \rangle$ and $F= \lambda_X \langle F_X \rangle$.
The 2 $\times$ 2 mass matrix for two neutral Higgs $H_u^0$ and $H_d^0$ is
\bea
\left( 
\begin{array}
{cc}
\mu^2  & B\mu \\
B\mu & \mu^2 
\end{array}
\right)
\left(
\begin{array}
{c}
H_u^0 \\
H_d^0
\end{array}
\right) .
\eea

Note that $|B\mu| = \mu^2$ holds at one loop if $\lambda_1=\lambda_2$.
In this limit, the CP even Higgs mass matrix is 
\bea
\mu^2 \left( 
\begin{array}
{cc}
1  & -1 \\
-1 & 1 
\end{array}
\right)
\left(
\begin{array}
{c}
H_u^0 \\
H_d^0
\end{array}
\right).\label{eq:Higgs_mass}
\eea
The determinant of the mass matrix is $0$
and the massless eigenstate $(H_u^0 +H_d^0)/\sqrt{2}$ appears
as a Goldstone boson.
Also there is a $SU(2)$ singlet Goldstone boson but it does not couple to the Higgs doublet directly and also they get the mass from loop corrections below the global symmetry breaking scale $f \sim M_N$.

There are three different scales $M_N$, $M$ and $\sqrt{F}$.
Both $X$ and $S$ couple to messengers and $M_N^2 \simeq \frac{F}{16\pi^2}$ is generated
at one loop from the K\"ahler potential ($X^\dagger S$) even if it was absent at tree level.

In the absence of the complete model,  the discussion is not valid since the top Yukawa coupling explicitly breaks the global symmetry. Therefore, the model needs the extra $T^\prime$
and $\bar{T}^\prime$ such that $Q, T^\prime$ can form a triplet under global $SU(3)$
and new Yukawa $y N_H T^\prime u^c$ should be present.
We assume that there are these partners with mass of order $f$ such that all the discussion of Goldstone boson Higgs can make sense.
It applies to other global symmetry examples. Later for the vector-like leptons, the presence of the extra lepton pair with mass of order $f$ is also assumed.

In summary the relevant scale is following.

\begin{itemize}
\item Messenger scale $M$ : can be heavy ($\ge f \sim \mu$)
\item Global symmetry breaking scale $f $ : TeV or heavy
\item Supersymmetric Higgs mass $\mu$ : TeV
\item Soft scalar mass $m_{\rm soft}$ : TeV
\end{itemize}

If the global symmetry is well preserved in the symmetry breaking sector and is only broken by the Yukawa couplings and gauge couplings at the scale $f \equiv \langle N \rangle$ ($=\frac{M_N}{\sqrt{\lambda_2}}$), the potential for the pseudo-Goldstone Higgs vanishes at the scale $f$,
\bea
V(h) & = & 0.\nn
\eea
Below the scale $f$, there would be a correction to the Higgs potential.
Indeed 
\bea
V(h) &\sim & - c(f,m_{\rm soft}) \frac{3}{8\pi^2} y_t^2 m_{\rm soft}^2 |h|^2 + \frac{3}{8\pi^2} y_t^4 \log (\frac{m^2_{\rm soft}}{m^2_{\rm top}}) |h|^4, 
\eea
where $c(f,m_{\rm soft})$ is a function of $f$ and $m_{\rm soft}$ and is  of order one when they are comparable, $f \sim m_{\rm soft}$.

In the MSSM, the light CP even Higgs gets a loop correction to the mass \cite{Okada:1990vk} \cite{Haber:1990aw} \cite{Ellis:1990nz}.
\bea
m_h^2 & \simeq & M_Z^2 + \frac{3y_t^2 m_t^2}{4\pi^2} \left[  \log \frac{m_{\tilde{t}}^2}{m_t^2} + \frac{X_t^2}{m_{\tilde{t}}^2} ( 1 - \frac{1}{12}  \frac{X_t^2}{m_{\tilde{t}}^2})  \right], 
\eea
where $X_t = A_t - \mu/\tan \beta$.

The physical Higgs mass, $m_h^2 \simeq {(125 \ {\rm GeV})}^2 \sim 2M_Z^2$, gives a picture on the size of the corrections to explain the observed Higgs mass for the Goldstone boson Higgs.
In the MSSM, tree level mass $m_h^2$ is at around $M_Z^2$
and one loop correction from top loop is about $M_Z^2$.

For $\tan \beta =1$, the tree level mass is zero as it is in the vanishing D-flat direction
and top/stop contribution is $\delta m_h^2 \sim M_Z^2$.
If there is a vector-like states and if there is an order one Yukawa couplings,
it would be possible to get a correction of order $M_Z^2$ from the one-loop running
which is comparable to the one from the top/stop loop.

\section{Phenomenological model below the global symmetry breaking scale}

In this section, we present a phenomenological model which is relevant for LHC measurement of Higgs and electroweak precision measurements.
As we have discussed in the previous section, we would not provide a complete model for global symmetry breaking but a phenomenological model of Psedudo-Goldstone Boson (PGB) Higgs with vector-like matter superfields.

In the context of extended MSSM, implications of vector-like matter have been discussed in literatures \cite{Moroi:1991mg, Moroi:1992zk, Maekawa:1995ha, Babu:2008ge, Martin:2009bg, Graham:2009gy, Moroi:2011aa, Martin:2012dg}.
Without introducing additional gauge symmetry, we would consider SM-like vector-like superfields that can couple to SM Higgs and SM superfields.
Such vector-like superfields with charge assignment  under $SU(3)_c\times SU(2)_L\times U(1)_Y$ are given by
\begin{eqnarray}
&&Q=({\bf 3},{\bf 2},1/6),\qquad\overline{Q}=({\bf \overline{3}},{\bf 2},-1/6),\\
&&U=({\bf 3},{\bf 1},2/3),\qquad\overline{U}=({\bf \overline{3}},{\bf 1},-2/3),\\
&&E=({\bf 1},{\bf 1},-1), \qquad\overline{E}=({\bf 1},{\bf 1},1).
\end{eqnarray}
which can be embedded in $\bf{10}+\bf{\overline{10}}$ under $SU(5)_G$, and
\begin{eqnarray}
&&L=({\bf{1}},{\bf{2}},-1/2),\qquad\overline{L}=({\bf{1}},{\bf{2}},1/2),\\
&&N=({\bf{1}},{\bf{1}},0),\qquad\overline{N}=({\bf{1}},{\bf{1}},0),\\
&&D=({\bf{3}},{\bf{1}},-1/3),\qquad\overline{D}=({\bf{\overline{3}}},{\bf{1}},1/3).
\end{eqnarray}
which can be embedded in $\bf{5}+\bf{\overline{5}}$.
For the gauge coupling unification, we should include the fields that form complete representations of $SU(5)$, so we can include $\bf{10}+\bf{\overline{10}}$, $\bf{5}+\bf{\overline{5}}$, or both.

From the recent measurement of Higgs decay \cite{:2012gk, :2012gu}, we have seen some substantial excess in diphoton channel but not in other channels
though the error bar is large at this stage.
If we anticipate that the diphoton excess compared to the Standard Model prediction survives even after the error bar is reduced,
it means that Higgs to di-photon coupling ($h-\gamma-\gamma$) is enhanced by some new physics effects but Higgs to gluon coupling ($h-g-g$) is not.
For such reason, we do not consider the vector-like particles with color since they also modify $h-g-g$ coupling.
By decoupling colored vector-like particles, we can construct a model with only vector-like leptons.
The superpotential is given by
\begin{equation}
W_{LNE}=M_L L\overline{L} +M_E E\overline{E} +M_N N\overline{N} + \hat{k}_EH_u\overline{L}E-\hat{h}_EH_dL\overline{E} +\hat{k}_NH_u{L}\overline{N}-\hat{h}_NH_d\overline{L}{N},
\label{eq:superpotential}
\end{equation} 
where $H_u$ and $H_d$ are Higgs superfields of MSSM. 

As discussed in the previous section, we consider TeV scale $\mu$ and $B\mu$ term which are related to global symmetry breaking scale.
Therefore, at that scale, CP-even Higgs mass matrix is given by (\ref{eq:Higgs_mass}), and one of the eigenstates is massless and is identified as SM Higgs.
This can be interpreted as MSSM light CP-even Higgs state for $\tan\beta=1$ and $m_A^2\sim B\mu\sim \mu^2$.
In the CP-even Higgs mixing relation,
\begin{equation}
\begin{pmatrix}
H \\ h
\end{pmatrix}
=
\begin{pmatrix}
\cos\alpha & \sin\alpha \\ -\sin\alpha & \cos\alpha
\end{pmatrix}
\begin{pmatrix}
h_d \\ h_u
\end{pmatrix},
\end{equation}
we can take $\alpha=\beta-\pi/2=-\pi/4$ and $h=(h_d +h_u)/\sqrt{2}$ as seen in the previous section.
The heavier eigenstate $H$ gets its mass of the order of $\mu\sim f$ and is integrated out below the global symmetry scale.
Fermionic components of Higgs doublets, Higgsinos are also integrated out at the same scale since $\mu$ is the Higgsino mass. 
Hence, our setup is effectively SM Higgs boson with matter superfields, i.e., we include whole supermultiplets for matter fields but we include only the lightest scalar component for Higgs supermultiplet.
Due to the renormalisation group (RG) running from the scale of global symmetry breaking to weak scale, quadratic and quartic potential of Higgs are developed as we discussed in the previous section.
In the following section, nevertheless, we keep the notation of $H_u$, $H_d$ and their VEVs $v_u$, $v_d$ for convenience.
For the physical observables, we will take $\tan\beta=1$.

In order to modify Higgs decay to di-photon, it is enough to introduce charge particle Yukawa couplings, $\hat{k}_E$ and $\hat{h}_E$ and the large Yukawa coupling can contribute Higgs quartic coupling for 125 GeV Higgs mass.
Then, we might decouple neutral particle by putting $M_N$ very high or $\hat{k}_N$ and $\hat{h}_N$ very small.
In such case, however, isospin symmetry is maximally broken and large Yukawa coupling $k_E$ and $h_E$ that are required to make Higgs mass 125 GeV substantially contribute oblique parameters $S$ and $T$.
If ($\hat{k}_N$, $\hat{h}_N$, $M_N$) are similar to ($\hat{k}_E$, $\hat{h}_E$, $M_E$), i.e., neutral sector are similar to charged sector, then isospin symmetry is approximately preserved so $T$ is highly suppressed. We will discuss this issue later.

In addition to oblique corrections, including neutral particles has another advantage.
Higgs mass can be generated partly by stop loop and partly by vector-like matter loop.  
Under the assumption that soft mass scale of vector-like particles is not much higher than TeV scale, it is somewhat difficult for loop correction of vector-like scalar to be as large as stop loop correction since leptonic particles do not have color factor 3.
On the other hand, if we introduce neutral vector-like particles which have similar Yukawa coupling as charged particles, they take part of sizable corrections for Higgs mass, i.e., it makes the effect of 2 copies of lepton pair.
This will be discussed in the following sections.

\section{Higgs mass in supersymmetry with vector-like states}

In calculating Higgs mass correction in the MSSM, it was found that one loop correction is positive and two loop correction is negative. 
Although two loop correction is smaller than one loop, it requires much heavier stop much than one loop case since the slope in the Higgs-stop mass plot is very mild at 125 GeV Higgs mass. However if one includes three loop correction, then it is not very different from one loop result \cite{Kant:2010}.

As the three loop result is more close to the one loop result rather than two loop computation, we calculate Higgs mass correction from extra vector-like matters only in one loop accuracy. We use one loop effective potential method. We consider $\tan{\beta}=1$ from now on.

\begin{figure}[t]
\includegraphics[width=0.6\textwidth]{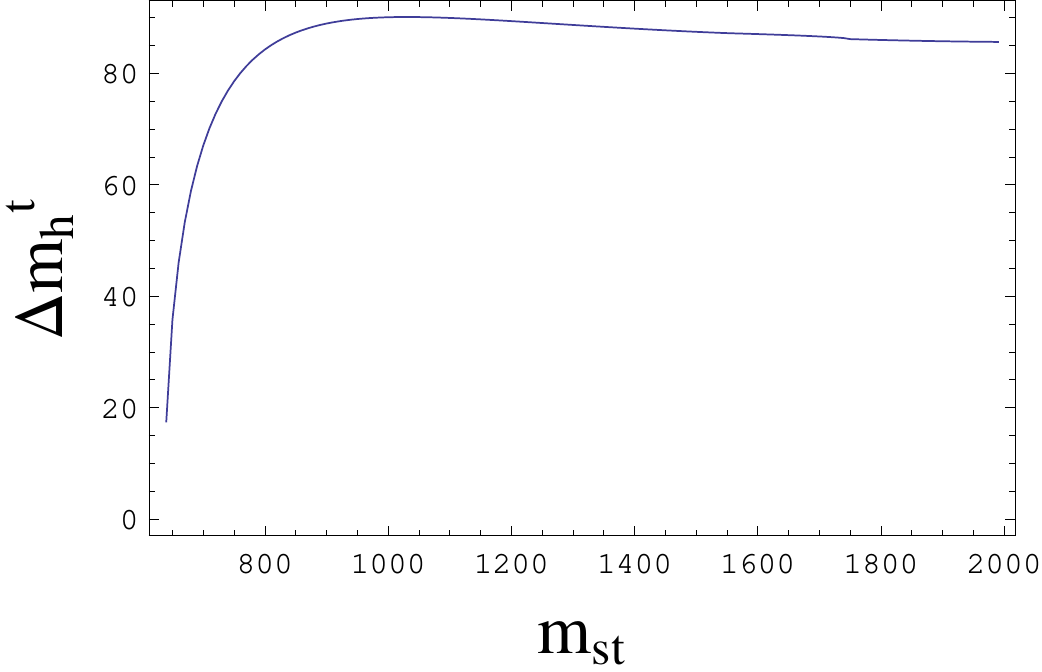}
\caption{FeynHiggs result of $\Delta m_h$ for MSSM contribution. $\mu$ is fixed by 2 TeV.}
\label{Fig0}
\end{figure}

Let us begin with the familiar top/stop contribution to the Higgs mass in the MSSM.
We assume that $\mu$ is order of $2 \sim 3$ TeV. 
Then the finite threshold correction from stop to Higgs mass is maximized 
when the stop mass is around $(2 \sim 3\text{ TeV})/ \sqrt{6} \simeq 0.8 \sim 1.2$ TeV without considering two loop effect.
Logarithim correction is larger 
for heavier stop but stop gives the largest finite threshold correction for the maximal stop mixing, $\sqrt{6} m_{\tilde{t}} \sim \mu$.
The FeynHiggs result is given by Fig.\ref{Fig0}. It is possible to make $\Delta (m_h^2)_{MSSM}$ to be order of $M_Z^2$ with $m_{\tilde{t}}\simeq 1$ TeV. 

In the MSSM part, our FeynHiggs result cannot be so accurate because extra vector-like states can modify it through 2-loop beta function. However, we assume that they do not have color charge and thus this effect is expected to be small enough.

The Fig. \ref{Fig0} shows the $\mu$ dependence of the Higgs mass correction for fixed stop mass. For $m_{\tilde{t}}<800$ GeV, $\Delta m_h^t$ rapidly increases because the non-logarithmic part is proportional to $-\mu^4/m_{\tilde{t}}^4$. And it gets local maximum for $\mu^2/m_{\tilde{t}}^2\simeq 2$. And there is an almost flat region since logarithmic part cancels the decreasing effect of finite threshold correction.

Now, we would proceed to analyze the vector-like matter effects on Higgs mass correction.
First, we would consider the  Higgs mass correction from charged fields, ($L$,$E$,$\overline{L}$,$\overline{E}$). The structure of the Higgs mass correction from neutral fields, ($L$,$N$,$\overline{L}$,$\overline{N}$) are exactly the same as charged part, so we can simply add for the real calculation. 
In order to calculate Higgs mass correction, we should identify masses for fermion components and scalar components of vector-like superfields.
From the superpotential (\ref{eq:superpotential}), we can easily read off the fermion mass matrix, which is given by
\begin{equation}
\begin{pmatrix}
E_L^c & E_R^c
\end{pmatrix}
{\cal M}_f
\begin{pmatrix}
E_L\\E_R
\end{pmatrix}
=
\begin{pmatrix}
E_L^c & E_R^c
\end{pmatrix}
\begin{pmatrix}
M_L & {k}_Ev\\ {h}_Ev & M_E
\end{pmatrix}
\begin{pmatrix}
E_L\\E_R
\end{pmatrix},
\end{equation}
where we redefine the charged fermion,
\begin{eqnarray}
&&L\equiv
\begin{pmatrix}
N_L\\E_L
\end{pmatrix},
\qquad
\overline{E}\equiv E_R^c,\\
&&\overline{L}\equiv
\begin{pmatrix}
E_L^c\\N_L^c
\end{pmatrix},
\qquad
E\equiv E_R.
\end{eqnarray}
Note that superscript $c$ does not mean charge conjugation.
We obtain
\begin{equation}
{\cal M}_f^{\dagger}{\cal M}_f=
\begin{pmatrix}
M_L^2+{h}_E^2v^2 & M_L{k}_Ev+M_E{h}_Ev \\
M_L{k}_Ev+M_E{h}_Ev & M_E^2+{k}_E^2v^2
\end{pmatrix},\label{mass:fermion}
\end{equation}
where we assume the masses and Yukawa couplings are real for simplicity. 
Together with superpotential (\ref{eq:superpotential}), soft SUSY breaking terms are also included for scalar components as follows.
\begin{equation}
\begin{split}
-{\cal L}_{\text{soft}}=&m_{{L}}^2|\widetilde{L}|^2+m_{{\overline{L}}}^2|\widetilde{\overline{L}}|^2+m_E^2|\widetilde{E}|^2+m_{\overline{E}}^2|\widetilde{\overline{E}}|^2\\
&+b_L\widetilde{L}\widetilde{\overline{L}}+b_E\widetilde{E}\widetilde{\overline{E}}+\hat{a}_{k}H_u\widetilde{\overline{L}}\widetilde{{E}}-\hat{a}_{h}H_d\widetilde{L}\widetilde{\overline{E}}+\text{h.c.}.
\end{split}
\end{equation}
The scalar mass matrix in basis $(\widetilde{E}_L^c,\widetilde{E}_R^c,\widetilde{E}_L^*,\widetilde{E}_R^*)$,
The scalar mass matrix is given by
\begin{equation}
{\cal M}_S^2=\overline{{\cal M}_f^2}+
\begin{pmatrix}
m_{\overline{L}}^2 & 0 & b_L^* & a_k^*v-{k}_E\mu v \\
0 & m_{\overline{E}}^2  & a_h^*v-{h}_E\mu v & b_E^* \\
b_L & a_hv-{h}_E\mu^*v & m_{L}^2 & 0 \\
a_kv-{k}_E\mu^*v & b_E & 0 & m_{E}^2 
\end{pmatrix}\label{mass:scalar}
\end{equation}
where 
\begin{equation}
\overline{{\cal M}_f^2}=
\begin{pmatrix}
{\cal M}_f{\cal M}_f^{\dagger} & 0 \\ 0 & {\cal M}_f^{\dagger}{\cal M}_f
\end{pmatrix}.\label{mass:fermion_sq}
\end{equation}

From the above mass matrices of fermions and scalars, we can calculate the one-loop effective potential, which is given by 
\begin{equation}
\Delta V = \frac {2N}{64\pi^2} \sum _{i=1}^{4} \biggl[ M_{S_i}^4 \biggl( \ln{\frac {M_{S_i}^2}{Q^2}}-\frac{3}{2} \biggr) -M_{F_i}^4 \biggl(\ln{\frac {M_{F_i}^2}{Q^2}}-\frac{3}{2}\biggr) \biggr],\label{CWpotential}
\end{equation}
where $M_{S_i}$ and $M_{F_i}$ are eigenvalues of mass matrices, (\ref{mass:scalar}) and (\ref{mass:fermion_sq}), respectively, and the whole expression is multiplied by $N$, the copy of vector-like matters.
From this effective potential, we can read off the Higgs quartic coupling,
\begin{equation}
\frac{\Delta\lambda}{4}h^4=\frac{1}{4!}\frac{\partial^4\Delta V(h)}{\partial h^4}\biggl|_{h=0}h^4,
\end{equation}
and 
\begin{equation}
\Delta m_h^2=2\Delta\lambda \langle h\rangle^2,
\end{equation}
where $\langle h\rangle=\sqrt{2}v$ and $v=174$ GeV.
We would consider a limiting situation so that a simple approximate expression can be obtained.
Let us take a limit of $M_L=M_E\equiv M_F$, $m_L^2=m_{\overline{L}}^2=m_{E}^2=m_{\overline{E}}^2\equiv m_s^2$ and $h_E=k_E$.
We also neglect $a_k$ and $a_h$ because of large $\mu$. However, it can be easily restored if we replace $\mu$ by $\mu-a/k$.
The eigenvalues of the vector-like fermions and scalars are given by 
\begin{eqnarray}
(M_F^2)_{\pm}&=&M_F^2+k_E^2v^2 \pm 2 k_E M_f v=(M_F\pm k_E v)^2,\\
(M_S^2)_i&=&\left\{ \begin{array}{ll} m_s^2+b+M_F^2+k_E^2v^2-2k_E v M_F+k_E\mu v \\m_s^2+b+M_F^2+k_E^2v^2+2k_E v M_F-k_E\mu v \\m_s^2-b+M_F^2+k_E^2v^2+2k_E v M_F+k_E\mu v \\m_s^2-b+M_F^2+k_E^2v^2-2k_E v M_F-k_E\mu v \end{array} \right .
.
\end{eqnarray}
Two scalar masses becomes smaller as b increases. At the Lagrangian level, it seems to be natural that b is order of $m_s M_F\simeq m_s^2/\text{(few)}$. However, in order not to have tachyonic scalar with large $\mu$,  $b$ much smaller than $m_s^2$ is considered. Thus from now on, consider $b$ is a small parameter. Also, $\mu$ cannot be larger than $(m_s^2+M_F^2)/k_E v$ for the same reason.
Neglecting $b$, the scalar mass can be rewritten as
\bea
M_S^2 = m_s^2 + (M_F^2)_{\pm} \pm \mu k_E v,
\eea
where $\pm$ means all four possibilities.

Plugging the above eigenvalues into Eq. (\ref{CWpotential}) and combining the neutral particle contribution, the one loop correction to the Higgs mass is given by
\bea
\Delta(m_{h}^2)^{\text{(vec)}}
&=& \frac{2N v^2(k_E^4+k_N^4)}{4\pi^2} \left[ \ln{x} + f(x) \right] + \Delta{m_h^2}_b,
\eea
where
\bea
x & = & \frac{(M_F^2+m_s^2)}{M_F^2},
\eea
and
\bea
f(x)&=& -\frac{1}{12} \left[ \frac{\mu^4}{(M_F^2+m_s^2)^2} -24(\frac{1}{2} -\frac{1}{x})\frac{\mu^2}{M_F^2+m_s^2}+ 16(1-\frac{1}{x})(2-\frac{1}{x})\right], \\
& = &  -\frac{1}{12} \left[ \frac{\mu^2}{(M_F^2+m_s^2)}-12(\frac{1}{2}-\frac{1}{x})\right]^2+\frac{32}{3}\left[(\frac{1}{x}-\frac{3}{8})^2-\frac{7}{64} \right].
\eea

\begin{figure}[t]
\includegraphics[width=0.48\textwidth]{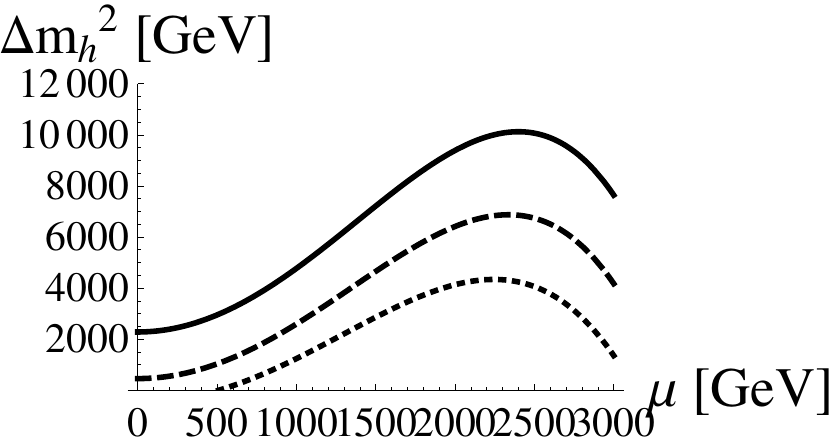}
\hfill
\includegraphics[width=0.48\textwidth]{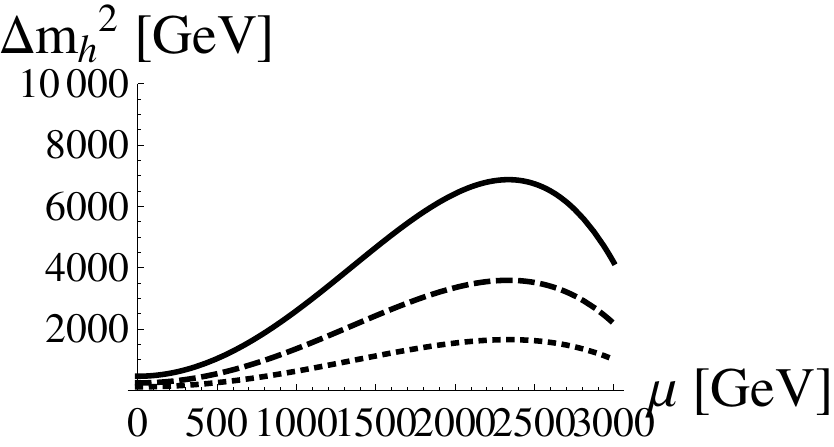}
\caption{Left plot is obtained by fixing $m_s=1$ TeV, $k_E=k_N=1.0$ and $M_F=(200, 300, 400)$ GeV for (thick, dashed, dotted) lines. Right plot is obtained by fixing $m_s=1$ TeV, $M_F=300$ GeV and $k_E=k_N=(1, 0.85, 0.7)$ for (thick, dashed, dotted) lines. The number of copy is one (N=1) for all cases.
\label{Fig1}}
\end{figure}

The function $f(x)$ represents the finite threshold correction when the vector-like scalar particles are integrated out. For $h_E=k_E$ and $h_N=k_N$, the finite threshold correction can not be large.
In the above expression, the first line shows that $f(x)$ is negative for $x \le 2$
as all three terms are negative. In order to obtain a positive threshold correction,
$x > 2$ is necessary. We restrict our discussion to the case of $x >2$, i.e., $m_s > M_F$.
Unlike the stop case, the finite threshold correction $f(x)$ has a maximum value $1/3$
which is very small compared to the stop case in which the finite threshold correction
is comparable to logarithmic contribution for $m_{\tilde{t}} \sim 1$ TeV.

If we take 
\bea
\mu^2 & = & 12(\frac{1}{2}-\frac{1}{x})(M_F^2+m_s^2)=6(m_s^2-M_F^2), 
\eea
the finite threshold correction to the Higgs mass is maximized. 
In the limit of large $x$ ($m_s \gg M_F$), the condition is the same as the stop case.
Therefore, if soft supersymmetry breaking scale is large compared to the vector-like fermion mass,
we can simultaneously maximize the finite threshold correction of stop and vector-like scalar by choosing the stop mass and the vector-like scalar mass to be the same.
More precisely, $m_s^2-M_F^2 = m_{\tilde{t}}^2$ is the best choice.
Fig. \ref{Fig1} shows the correction to the Higgs mass as a function of $\mu$ for given $m_s$. It also shows the dependence on $M_F$ and $k$, respectively. 
$\Delta{m_h^2}$ is proportional to $k^4$. Thus 1.2 times large $k$ provides twice larger correction.

\begin{figure}[t]
\centering
\includegraphics[width=0.48\textwidth]{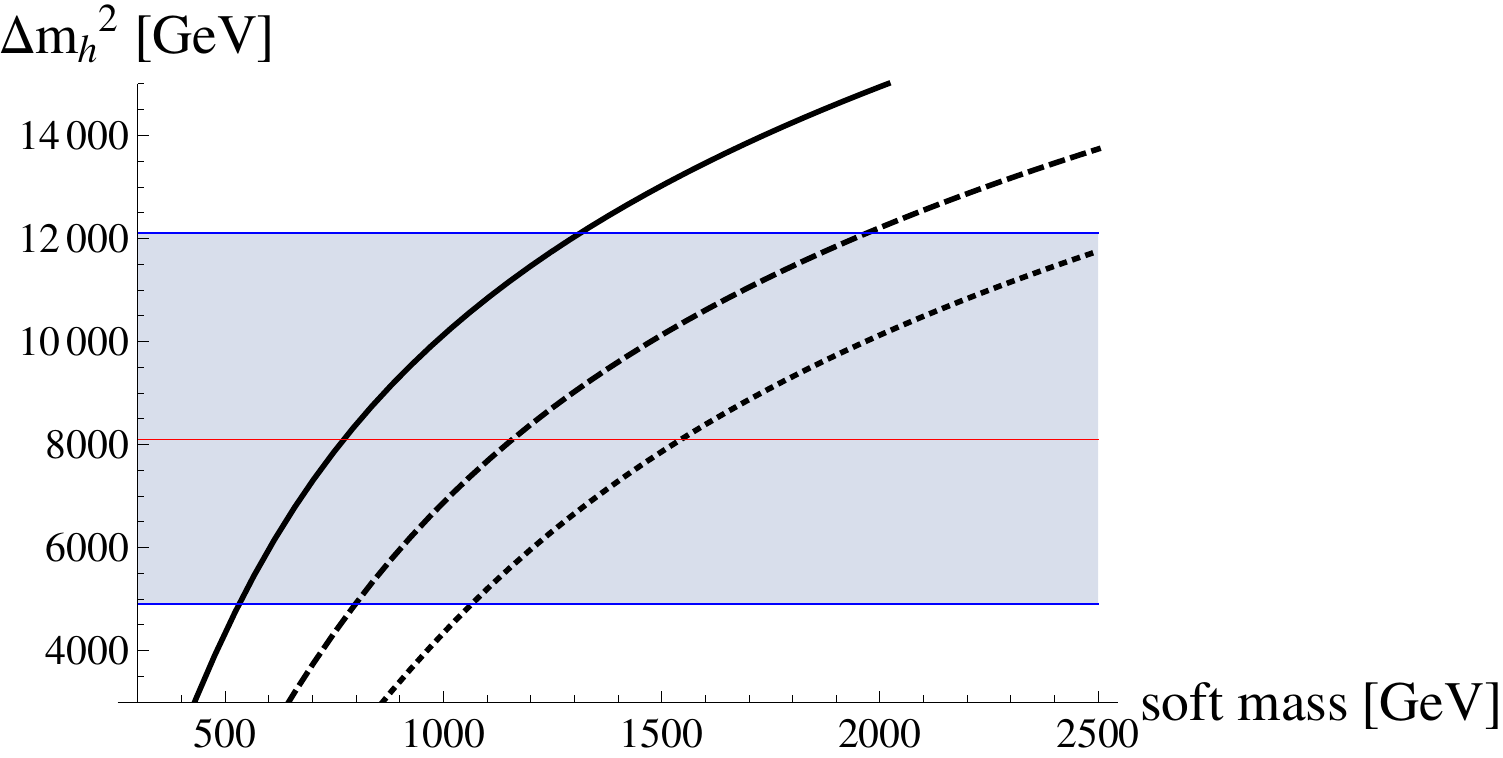}
\hfill
\includegraphics[width=0.48\textwidth]{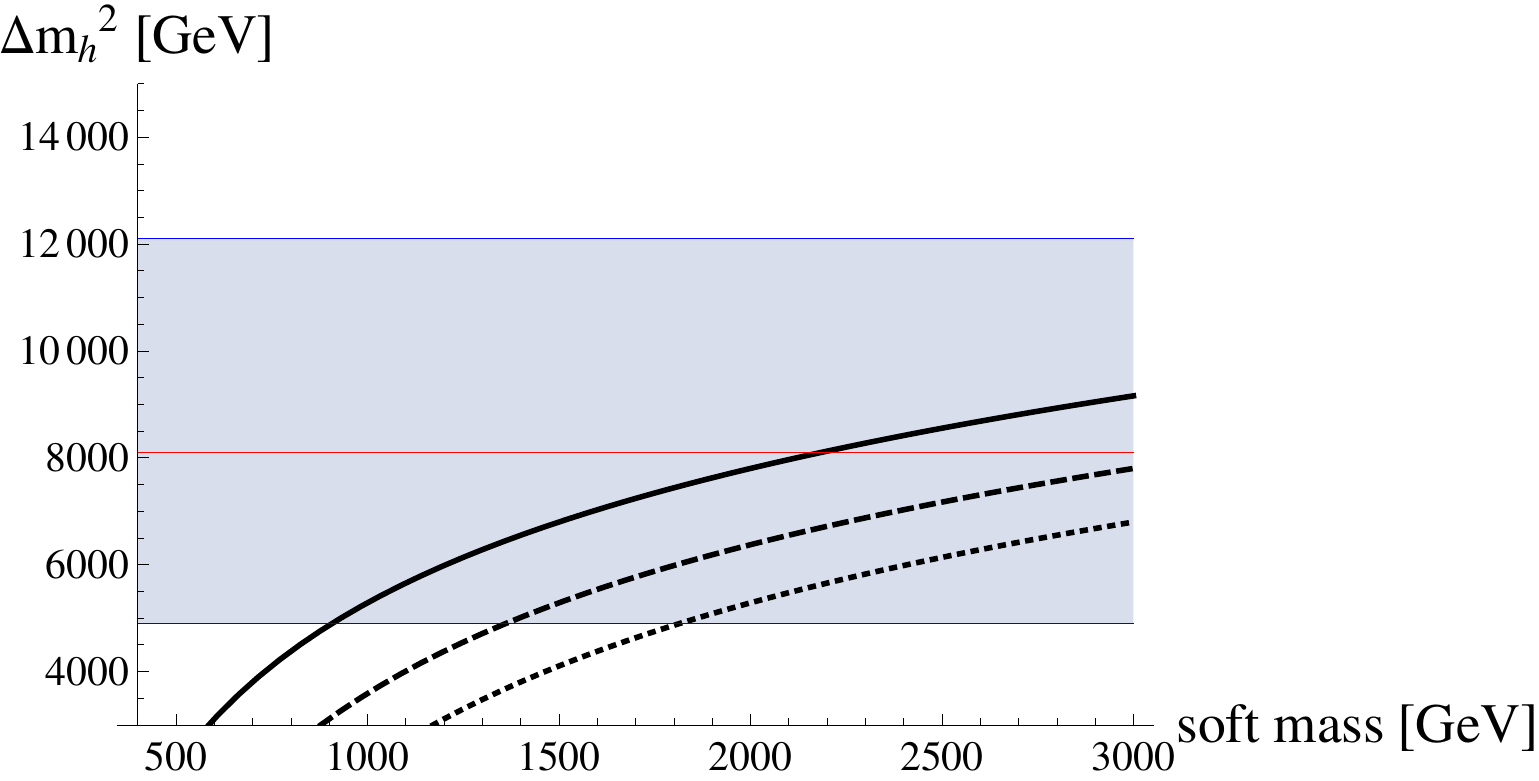}
\caption{Maximum value of $m_h^2$ for given $M_F$ with N=1. The (Thick, Dashed, Dotted) lines correspond to $M_F=(200, 300, 400)$ GeV. Red line is $m_h^2=(90\text{ GeV})^2$ and blue lines are $\Delta m_h^2=((90\pm20) {\ \rm GeV})^2$. The left is for $k_E=k_N=1.0$ and the right is for $k_E=k_N=0.85$}
\label{Fig2}
\end{figure}

\begin{figure}
\includegraphics[width=0.49\textwidth]{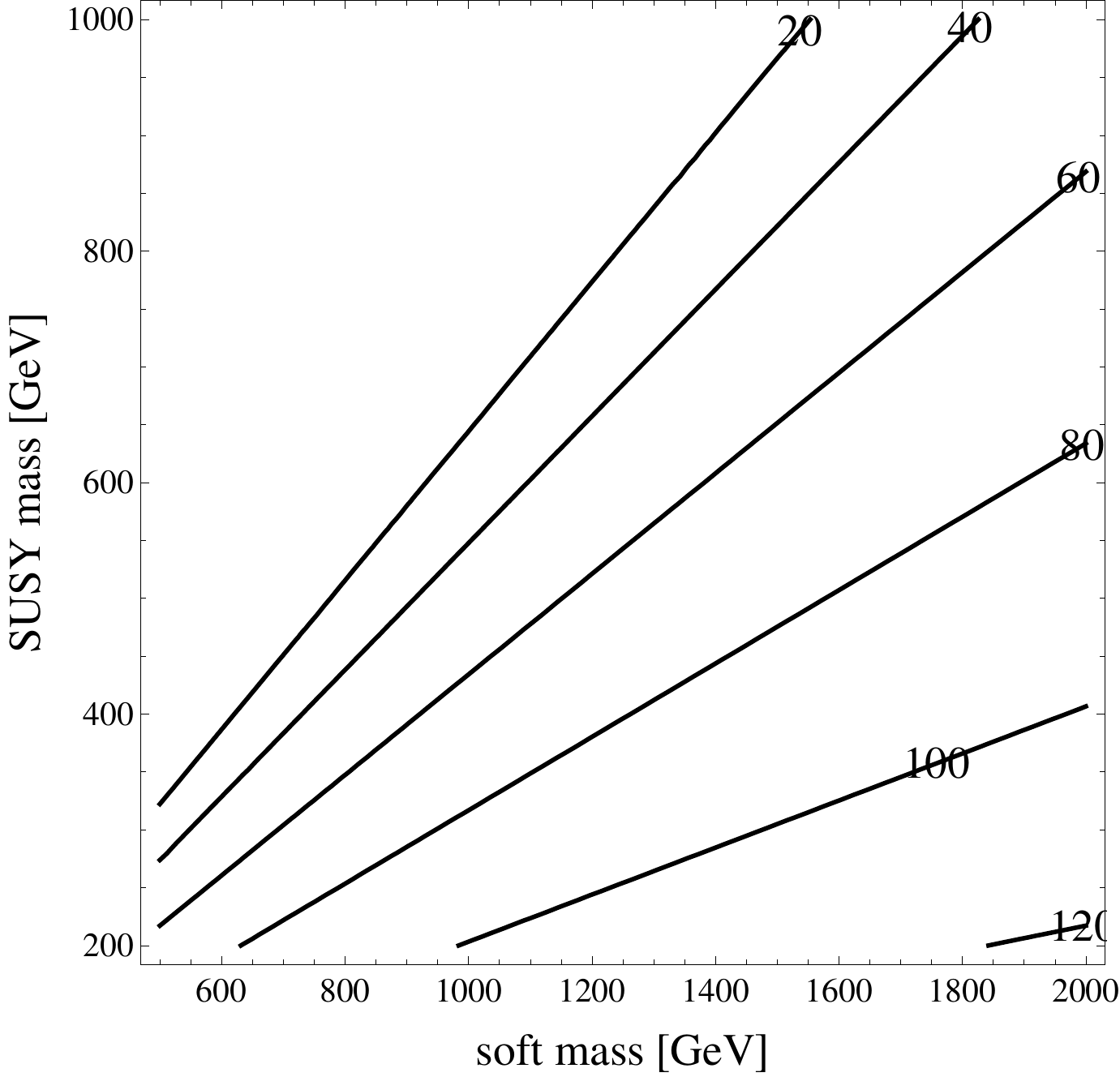}
\hfill
\includegraphics[width=0.49\textwidth]{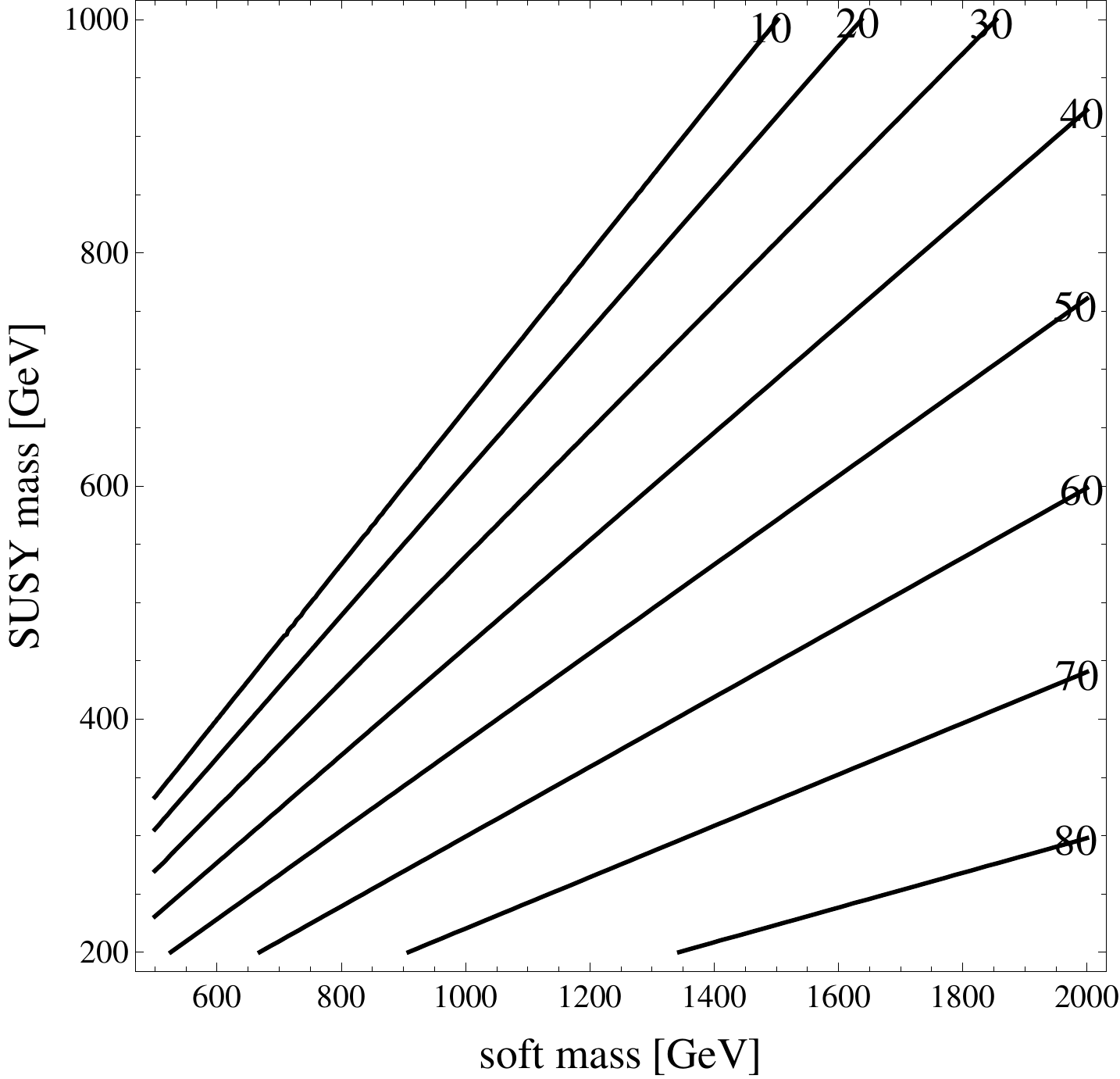}
\caption{Maximum value of $\sqrt{(\Delta m_h^2)^{\text{(vec)}}}$ for each $m_s$ and $M_f$ with $N=1$. The left is for $k_E=k_N=1.0$ and the right is for $k_E=k_N=0.85$}
\label{Fig3}
\end{figure}

These values are plotted in Fig.\ref{Fig2} and Fig.\ref{Fig3}. From Fig.\ref{Fig2}, we know that it is possible to make $\Delta (m_h^2)^{\text{(vec)}}\simeq M_Z^2$ with one copy of vector-like matters if $k=1.0$ and $m_s$  is around 1 TeV. Two copies are needed for $k=0.85$.

Even for large $\mu$, the scalar mass squared should be positive in order not to break $U(1)_{\rm em}$. 
The lightest scalar mass eigenvalue is given by 
\begin{eqnarray}
& &m_s^2+M_F^2+k^2v^2-2k v M_F-k\mu v\nn\\
&=&m_s^2+M_F^2+k^2v^2-2k v M_f-\sqrt{6}k v \sqrt{m_s^2-M_F^2} \nn\\
&>&(M_f-k v)^2+m_s(m_s-\sqrt{6} k v)\nn\\
&>&m_s(m_s-\sqrt{6}k v). 
\end{eqnarray}
In the second line, the condition for the maximum correction to the Higgs mass is used.
If $m_s>\sqrt{6}k v\simeq 426$ GeV, the scalar mass squared still remains to be positive
for the choice of $\mu$ which gives the maximum threshold correction to the Higgs mass
and there is no problem of $U(1)_{\rm em}$ breaking.

The correction proportional to the $b$ parameter is,
\bea 
\Delta(m_{h}^2)_b & \simeq & -\frac{2N v^2(k_E^4+k_N^4)}{3\pi^2}\frac{\mu M_F (\mu^2+M_F^2-3m_s^2)}{(M_F^2+m_s^2)^3}b.
\eea
We also take $b_N=b_E=b$. This agrees with the result in \cite{Martin:2009bg} setting $k=h$ and $\tan{\beta}=1$. The effect from $b$ is suppressed to be
\bea
-\frac{2\sqrt{3}N v^2 (k_E^4+k_N^4)}{ \pi^2} \frac{M_F}{\sqrt{m_s^2 + M_F^2}} \frac{b}{m_s^2+M_F^2}, \nn
\eea
for $\mu \sim \sqrt{6}m_s$ and very large $m_s$. Thus we also neglect $\Delta(m_{h}^2)_b$ from now on.

\section{Higgs decay to diphoton with vector-like fermions}

In this section, we consider the modification of Higgs decay to photon pair.
The partial decay width of diphoton decay is given by
\begin{equation}
\Gamma(h\to\gamma\gamma)=\frac{\alpha^2m_h^3}{1024\pi^3}
\biggl|
\frac{g_{hVV}}{m_V^2}Q_V^2A_1(\tau_V)+\frac{2g_{hff}}{m_f}N_{c,f}Q_f^2A_{1/2}(\tau_f)+N_{c,S}Q_S^2\frac{g_{hSS}}{m_S^2}A_0(\tau_S)
\biggr|^2,
\end{equation}
where $V$, $f$ and $S$ stand for vector, fermion and scalar particles.
$Q_i$'s are $i$th particles' electromagnetic charge and $N_{c,f(S)}$ is internal degree of freedom such as color.
$g_{hii}$ can be obtained by the mass eigenvalue of each particle by taking $g_{hii}=\partial m_i/\partial h$ for fermions and $g_{hii}=\partial m_i^2/\partial h$ for scalars and vector bosons.
Note that $\langle h\rangle=\sqrt{2}v$ and $v=174$ GeV as before.
Loop functions $A_i(\tau)$'s are given in Appendix and $\tau_i=4m_i^2/m_h^2$.
In the SM, the dominant contribution is from $W$ boson loop, and the subdominant one is from top quark loop.
For 125 GeV Higgs, the numerical values are given by
\begin{equation}
A_1(\tau_W)=-8.32,\qquad N_cQ_t^2A_{1/2}(\tau_t)=1.84.\label{amp:SM}
\end{equation}
Top quark contribution destructively interfere the dominant $W$ boson contribution.
In the case of chiral fermion that couple to Higgs as SM fermions, the situation is the same as in the top quark case.
According to \cite{Carena:2012xa}, large mass mixing between chiral fermions states  that is induced by Yukawa coupling can make the interference constructive, i.e. $g_{hff}$ can be negative after diagonalization.
Scalar contribution is also similar to fermion case, large mixing between scalars can make one of  $g_{hSS}$ negative.

In order to easily see the condition for constructive interference, It is  illustrative to consider the limit that the charged particles in the loop are very heavy limit.
The effective lagrangian for Higgs-photon-photon coupling is derived by low energy Higgs theorem \cite{Ellis:1975ap, Shifman:1979eb}.
The effective lagrangian is given by
\begin{equation}
{\cal L}_{h\gamma\gamma}=\frac{\alpha F_{\mu\nu}F^{\mu\nu}}{16\pi}\frac{h}{\sqrt{2}v}\biggl[
\sum_i b_i\frac{\partial}{\partial \log v}\log\big(\det{\cal M}^{\dagger}_{F,i}{\cal M}_{F,i}\big)+\sum_i b_i\frac{\partial}{\partial \log v}\log\big(\det{\cal M}^{2}_{B,i}\big)
\biggr]\label{eq:hgamgam}
\end{equation}
where ${\cal M}_{F,i}$ and ${\cal M}_{B,i}$ are the mass matrices of fermions and bosons.
$b_i$'s are the beta function coefficients which are given by
\footnote{Higgs-gluon-gluon coupling can be obtained by replacing $2N_{c,f(s)}Q_{f(s)}^2\alpha$ with $\alpha_s\delta^{ab}$.}
\begin{eqnarray}
b_{1/2}&=&\frac43 N_{c,f}Q_f^2\quad\text{for a Dirac fermion},\\
b_1&=&-7\quad\text{for the W boson},\\
b_0&=&\frac13N_{c,S}Q_S^2\quad\text{for a charged scalar}.
\end{eqnarray}
For LNE model, we only care about LE, because N doesn't change Higgs-photon-photon coupling. 
If the mass matrix is in ther form such as
\begin{equation}
{\cal M}_f^{\dagger}{\cal M}_f=
\begin{pmatrix}
m_{11}^2 & m_{12}^2 \\ m_{12}^{*2} & m_{22}^2
\end{pmatrix},
\end{equation}
we find
\begin{equation}
\frac{\partial}{\partial v}\log\big(\det{\cal M}_f^{\dagger}{\cal M}_f\big)
=\frac{1}{m_{11}^2m_{22}^2-|m_{12}^2|^2}
\biggl(
m_{11}^2\frac{\partial}{\partial v}m_{22}^2+m_{22}^2\frac{\partial}{\partial v}m_{11}^2-\frac{\partial}{\partial v}|m_{12}^2|^2
\biggr).
\label{eq:log_der}
\end{equation}
From the fermion mass matrix (\ref{mass:fermion}), we obtain
\begin{eqnarray}
&&\frac{\partial}{\partial v}\log\big(\det{\cal M}_f^{\dagger}{\cal M}_f\big)\nn\\
&&=\frac{2}{\det{\cal M}_f^{\dagger}{\cal M}_f}
\big\{
(M_L^2+{h}_E^2v^2){k}_E^2v+(M_E^2+k_E^2v^2)h_E^2v
-(M_Lk_E+M_Eh_E)^2v
\big\}\nn \\
&&=-\frac{4{k}_E{h}_E v}{M_E M_L -{k}_E{h}_E v^2},
\label{eq:log_der_ferm}
\end{eqnarray}
If $M_LM_E>{k}_E{h}_Ev^2$, the first (second) quantity become negative so that such contribution contructively interfere dominant $W$ boson loop.
Moreover, the amount of fermion contribution is proportional to product of Yukawa couplings, $h_Ek_E$, in the limit of large SUSY mass, $M_L$ and $M_E$. 
It agrees with the fact that large mixing between fermions make fermion loop contructively interfere $W$ boson loop.

In the similar way as fermion case, we can see the scalar contribution to Higgs-photon-photon coupling from the scalar mass matrix (\ref{mass:scalar}).
In order to visualize in easier way, we can change the basis of the scalar mass matrix as the following.
\begin{equation}
V^{\dagger}{\cal M}_S^2V=
\begin{pmatrix}
A & C \\ C^{\dagger} & B
\end{pmatrix},
\end{equation}
with
\begin{equation}
V=
\begin{pmatrix}
1 & 0 & 0 & 0 \\ 0 & 0 & 0 & 1 \\ 0 & 0 & 1 & 0 \\ 0 & 1 & 0 & 0
\end{pmatrix}.
\end{equation}
The $A$, $B$ and $C$ are $2\times2$ matrices which are given by
\bea
A&=&
\begin{pmatrix}
M_L^2+{k}_E^2v^2+m_{\overline{L}}^2 & a_kv-{k}_E\mu v\\
a_kv-{k}_E\mu v & M_E^2+{k}_E^2v^2+m_{E}^2
\end{pmatrix},
\eea
\bea
B&=&
\begin{pmatrix}
M_L^2+{h}_E^2v^2+m_{L}^2 & a_hv-{h}_E\mu v\\
a_hv-{h}_E\mu v & M_E^2+{h}_E^2v^2+m_{\overline{E}}^2
\end{pmatrix},
\eea
\bea
C&=&
\begin{pmatrix}
b_L & {h}_EM_Lv+{k}_EM_Ev \\ {k}_EM_Lv+{h}_EM_Ev & b_E
\end{pmatrix}.
\eea
Note that we assume all parameters are real.
In the limit of very small $C$, $A$ and $B$ can be separated into two independent $2\times2$ matrices and look very similar to ordinary slepton mass matrices.
In the limit of $\mu \sim m_S\gg M_L\sim M_E$ and $m_S^2\gg b$, $C$ is subdominant part of whole mass matrix so that we can deal with it as perturbation.
Moreover, its contribution is on rise in second order, i.e., ${\cal O}(v^2/\mu^2)$.
Thus, we can obtain the diagonal mass matrix at the leading order,
\begin{equation}
\begin{pmatrix}
U_1^{\dagger} & 0 \\ 0 & U_2^{\dagger}
\end{pmatrix}
\begin{pmatrix}
A & C \\ C^{\dagger} & B
\end{pmatrix}
\begin{pmatrix}
U_1 & 0 \\ 0 & U_2
\end{pmatrix}
=
\begin{pmatrix}
U_1^{\dagger}AU_1 & U_1^{\dagger}CU_2 \\ 
U_2^{\dagger}C^{\dagger}U_1 & U_2^{\dagger}BU_2
\end{pmatrix}
\end{equation}
where 
\begin{equation}
U_1=
\begin{pmatrix}
\cos\theta_1 & -\sin\theta_1 \\ \sin\theta_1 & \cos\theta_1
\end{pmatrix},\qquad
U_2=
\begin{pmatrix}
\cos\theta_2 & -\sin\theta_2 \\ \sin\theta_2 & \cos\theta_2
\end{pmatrix}
\end{equation}
and
\begin{eqnarray}
\tan2\theta_1&=&\frac{2(a_kv-{k}_E\mu v)}{M_L^2-M_E^2+m_{\overline{L}}^2-m_{E}^2},\\
\tan2\theta_2&=&\frac{2(a_hv-{h}_E\mu v)}{M_L^2-M_E^2+m_{{L}}^2-m_{\overline{E}}^2}.
\end{eqnarray}
Neglecting the off-block contribution from $C$ matrix which will appear in the sceond order of perturbation, we simply obtain the scalar contribution from Eq. (\ref{eq:log_der}) as in the fermion case.
It is given by
\begin{equation}
\begin{split}
\frac{\partial}{\partial v}\log\big(\det {\cal M}_S^2\big)\approx&\frac{2(M_L^2+m_{\overline{L}}^2+M_E^2+m_{E}^2){k}_E^2v-2{k}_E^2\mu^2 v}{(M_L^2+m_{\overline{L}}^2)(M_E^2+m_E^2)}\\
&+\frac{2(M_L^2+m_{{L}}^2+M_E^2+m_{\overline{E}}^2){h}_E^2v-2{h}_E^2\mu^2 v}{(M_L^2+m_{{L}}^2)(M_E^2+m_{\overline{E}}^2)}.
\end{split}
\label{eq:log_der_scal}
\end{equation}
In order to make constructive interference, we should satisfy $\mu^2>M_L^2+m_{\overline{L}}^2+M_E^2+m_{E}^2$ and $\mu^2>M_L^2+m_{{L}}^2+M_E^2+m_{\overline{E}}^2$.
However, since we consider the case of $\mu \sim m_S\gg M_L\sim M_E$ to make sizable Higgs mass correction, the effect is not large.

\begin{figure}[t]
\includegraphics[width=0.49\textwidth]{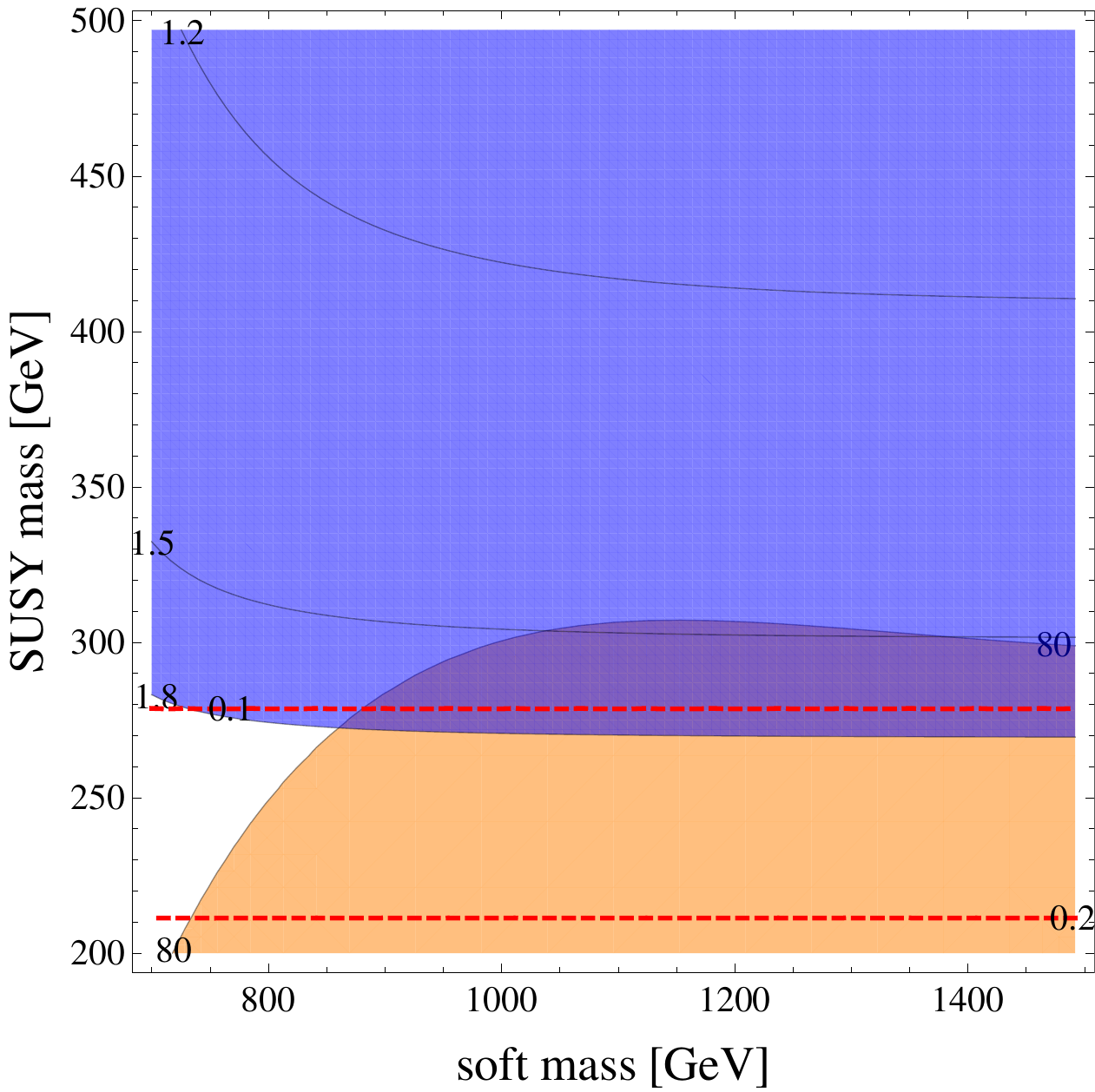}
\hfill
\includegraphics[width=0.49\textwidth]{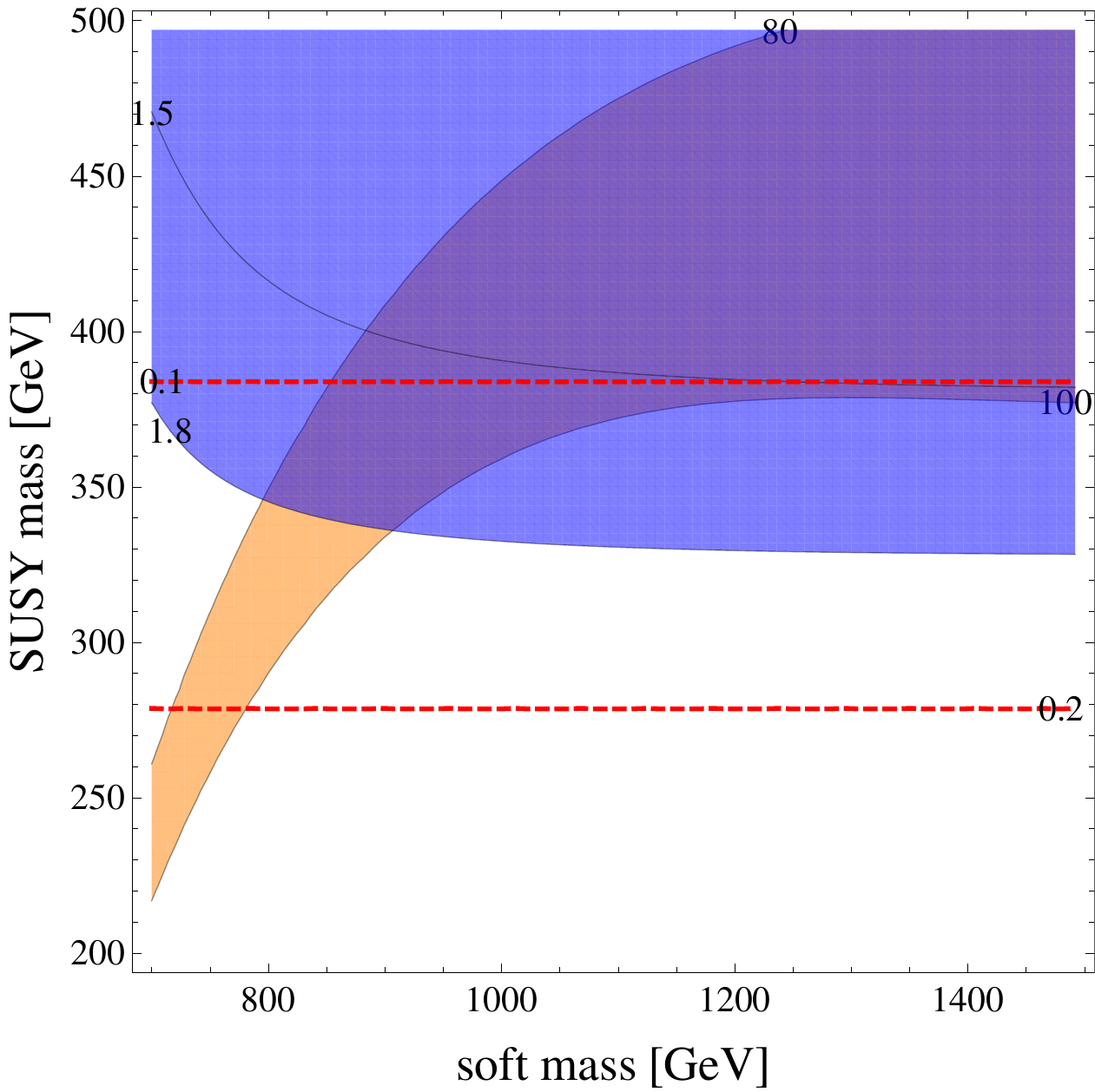}
\caption{Orange region is for $\sqrt{(\Delta m_h^2)^{\text{(vec)}}}=80$ - $100 $ GeV. Blue regions is $R_{\gamma \gamma}$ smaller than 1.8. Red dashed line denotes $\Delta S=0.1$ and $0.2$, respectively. $\mu$ is fixed by $2000$ GeV and (stop soft mass) $=1000$ GeV. Here, $k_N=k_E=1.0$. Left and right correspond to $N=1$ and $2$, respectively.
The light vector-like fermion mass can be read from the relation with the SUSY mass, $M_{F-}=M_F - k v$.
For $k=1$, $M_F - 174$ GeV is the light fermion mass.
}
\label{Fig4}
\end{figure}

\begin{figure}[t]
\includegraphics[width=0.49\textwidth]{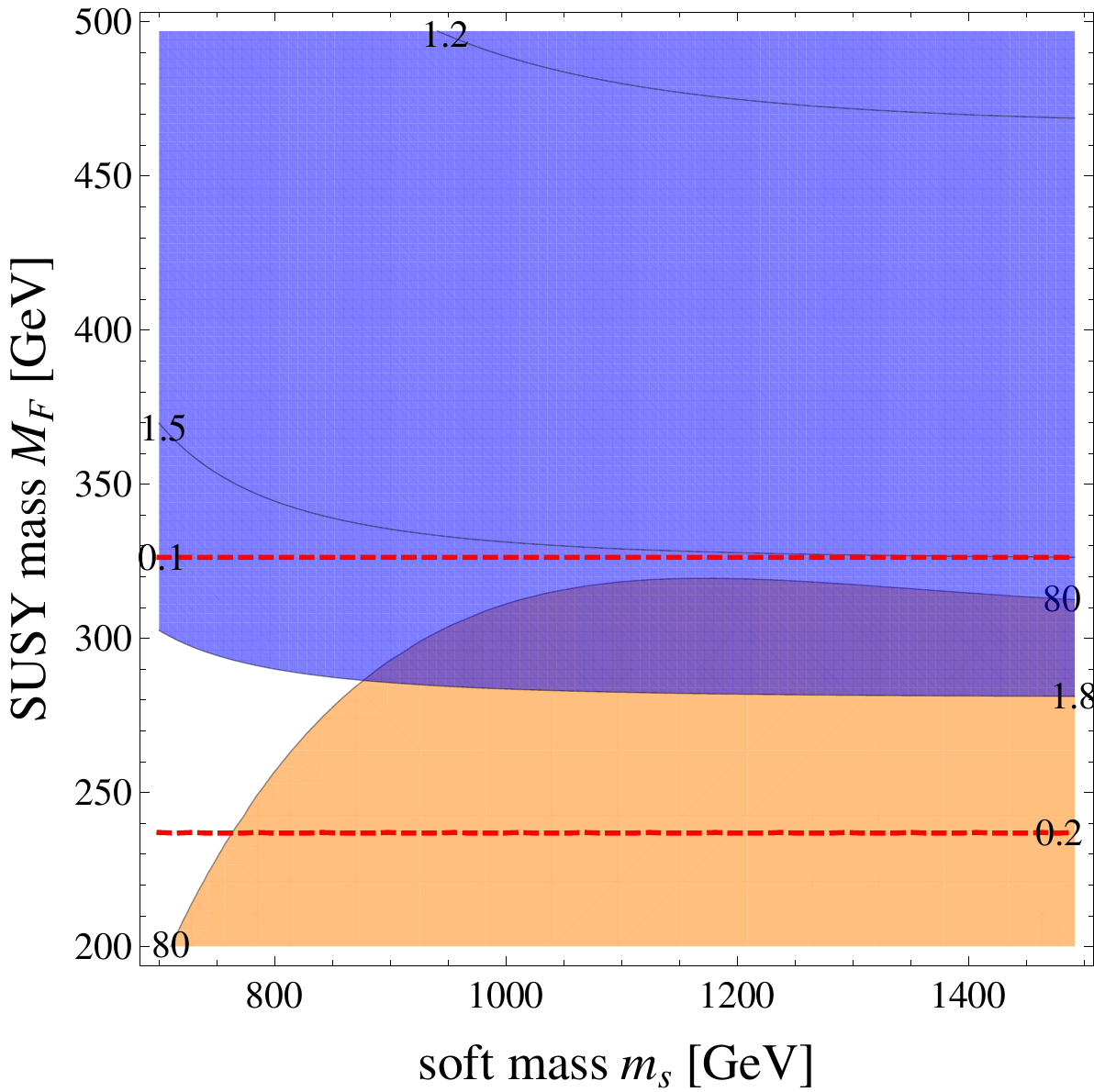}
\hfill
\includegraphics[width=0.49\textwidth]{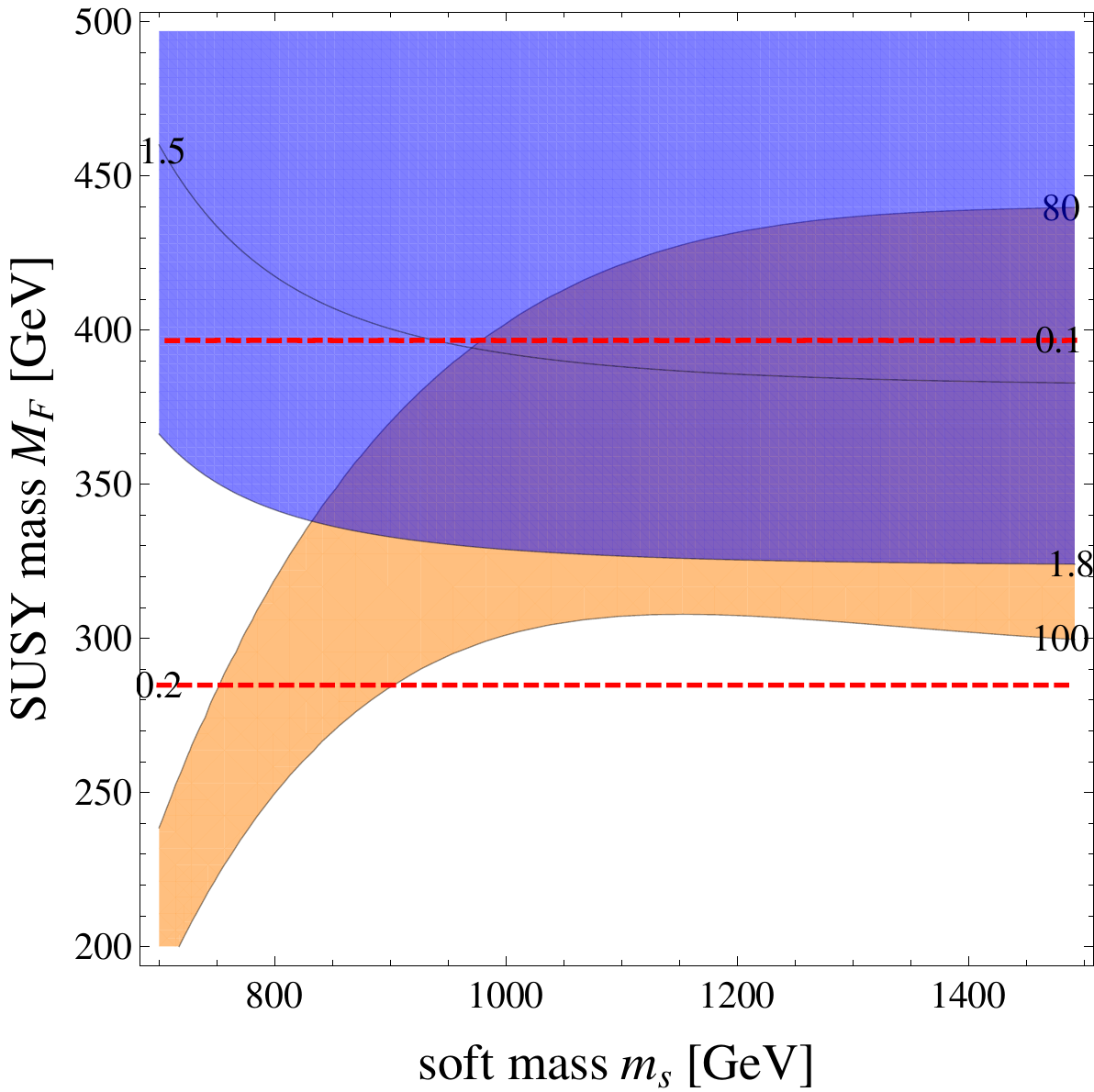}
\caption{Everything is the same with Fig. \ref{Fig4} except that $k_N=k_E=0.85$ and the left and right correspond to $N=2$ and $3$, respectively.
The light vector-like fermion mass can be read from the relation $M_{F-}=M_F - k v$.
For $k=0.85$, $M_F - 148$ GeV is the light fermion mass.
}
\label{Fig5}
\end{figure}

In such limit that we consider in this paper, we can easily use the above equations to estimate the enhancement effect for Higgs decay to diphoton since the loop functions rapidly approach to its asymptotic values once the mass of the particle is larger than Higgs mass.
For example, if the mass is around twice of Higgs mass, we can deal with this within 10\% error.
One more thing to emphasize is that the loop function of fermion and scalar decrease monotonically as the mass increases, so the result in such limit is generally smaller than real value.
Plugging Eqs. (\ref{eq:log_der_ferm}) and (\ref{eq:log_der_scal}) into (\ref{eq:hgamgam}), we obtain the approximate result in the leading order,
\begin{equation}
\begin{split}
{\cal L}^{\text{vec}}_{h\gamma\gamma}\approx&\frac{\alpha F_{\mu\nu}F^{\mu\nu}h}{16\sqrt{2}\pi}\biggl[
\frac43\biggl(-\frac{4k_Eh_Ev}{M_EM_L-k_Eh_Ev^2}\biggr)\\
&+\frac23\frac{(m_{\overline{L}}^2+m_E^2-\mu^2)k_E^2v}{m_{\overline{L}}^2m_E^2}
+\frac23\frac{(m_{{L}}^2+m_{\overline{E}}^2-\mu^2)h_E^2v}{m_{{L}}^2m_{\overline{E}}^2}
\biggr]\\
\approx&-\frac{\alpha F_{\mu\nu}F^{\mu\nu}h}{16\pi}\times\frac{8\sqrt{2}}{3} \frac{{h}_E{k_E}v}{M_L M_E-k_Eh_Ev^2}.
\end{split}
\end{equation}
From the above result, we can estimate the enhancement of partial decay width of Higgs to diphoton.
Using the SM values for $W$ boson and top quark in (\ref{amp:SM}), we can obtain simple relation,
\begin{equation}
\begin{split}
R_{\gamma\gamma}=&\biggl[1+\frac{{\cal A}_{\text{NP}}}{{\cal A}_{\text{SM}}}\biggr]^2\\
\simeq&\biggl[1+\biggl(\frac{v}{\sqrt{2}(-8.32+1.84)}\biggr)\biggl(\frac{8\sqrt{2}}{3} \frac{{h}_E{k_E}v}{M_L M_E-h_E k_E v^2}\biggr)\biggr]^2,
\end{split}
\end{equation}
where we neglect the stop contribution in the second line since stop masses are large to make one loop correction $M_Z^2$.
With MSSM stop contribution, Fig. \ref{Fig4} and Fig. \ref{Fig5} are obtained.
In all of the plots, the light stop contribution to $h-g-g$ coupling is 0.01 and $h-g-g$ coupling just like the Standard Model coupling.

The first parenthesis in the second line is the inverse of sum of $W$ and top contribution, while the second one is vector-like fermion contribution.
$R_{\gamma\gamma}=1.46$ for $M_L=M_E=300$ GeV, and $R_{\gamma\gamma}=1.20$ for $M_L=M_E=400$ GeV. These numbers coincide very well with the numerical result depicted in Fig. \ref{Fig4} when number of copy is 1 and soft mass is very large.

For $M_L M_E \gg k_E h_E v^2$, the expression is further simplified,
\bea
R_{\gamma\gamma}-1 & \simeq & 0.8 \frac{h_E k_E v^2}{M_L M_E} + {\cal O} ( h_E^2 K_E^2 v^4/M_L^2 M_E^2).
\eea

There is a correlation between the Higgs mass correction from the vector-like states
and the di-photon enhancement if soft scalar mass of the vector-like states is given.
For given number of copy $N$, Yukawa couplings $k_E=h_E=k$, and the vector-like scalar soft mass $m_s$, there is the upper bound on the vector-like fermion mass to give $\Delta m_h^2 = M_Z^2$.
Then in turn it gives a lower bound on $R_{\gamma\gamma}-1$, the minimum enhancement of the di-photon event rate.
In the limit of $m_s \gg M_F \gg kv$, the minimum enhancement of di-photon event has a simple analytic expression,
\bea
R_{\gamma \gamma} -1 &\gtrsim & 0.8 \frac{Nk^2 v^2}{m_s^2} \exp{\left(\frac{\pi^2 M_Z^2}{Nk^4 v^2}-\frac{1}{3}\right)}.
\eea
The minimum enhancement is obtained from the upper bound for $M_F$ needed to raise the Higgs mass for given $m_s$ (and $\mu$ for the maximal mixing). 
Of course the minimum enhancement rate disappears for large $m_s$ with appropriate choice of $\mu$ for the maximal mixing. For given $k$, there is an optimum choice of $N$ which makes the minimum of the $R_{\gamma \gamma} -1$ to be the smallest.
If the soft supersymmetry breaking parameters are constrained to be smaller than 1 TeV (1.4 TeV), the Higgs to di-photon rate is predicted to be enhanced at least by factor about 1.6 (1.3) for $k=1$ and $N=1$. If $N=2$ and $k=1$, the Higgs to di-photon rate has the minimum enhancement of 1.3 (1.2) for $m_s =1$ TeV ($1.4$ TeV). 

Finally we add a few comments on the effects of vector-like fermions on the electroweak precision observables. The extra vector-like doublets having Yukawa couplings with the Higgs
can make a contribution to $S$ and $T$ parameters.
Unless the vector-like fermion is extremely light, we can simplify the correction of the vector-like fermions to $S$ ($k v \ll M_F$). 
In general, vector-like fermion doublet $(U,D)$ contributes to $S$ and $T$.
If $M_U=M_D=M_F$, $h_U=k_U$, $h_D=k_D$, the expression can be made simple.
\bea
\Delta S & = & \frac{N}{6\pi} \left[ -2Y \log (\frac{m_{U_1}^2 m_{U_2}^2}{m_{D_1}^2 m_{D_2}^2}) \right] +
 \frac{11N}{30\pi} \left[ ( \frac{k_U v}{M_F} )^2 +  ( \frac{k_D v}{M_F} )^2 +{\cal O} (( \frac{k v}{M_F} )^4 ) \right].
\eea
Here $Y$ is the hypercharge of the doublet fermion and $m_{U_i}$ ($m_{D_i}$) is the eigenvalue of the upper (lower) component fermions with $i=1,2$ ($U=N$, $D=E$ in our case).
The first term can take any sign depending on the mass eigenstate
while the second term is always positive definite.
Therefore, there is a parameter space in which two contribution can cancel with each other.
In the special limit of degenerate spectrum for up and down fermions ($k_E=h_E$ in the $LND$ model),
the first term proportional to $Y$ disappears
and the contribution to $S$ is positive definite.
In the Fig. \ref{Fig4} and Fig. \ref{Fig5}, $\Delta S =0.1, 0.2$ contours are plotted.

If $M_U=M_D=M_F$, $h_U=k_U$, $h_D=k_D$, the expression for $T$ can also be made simple.
\bea
\Delta T & = & \frac{N}{10\pi \sin^2 \theta_W m_W^2} \left[ \frac{(k_U^2-k_D^2)^2 v^4}{M_F^2}
+ {\cal O} ( \frac{(k v)^6}{M_F^4}  ) \right].
\eea
The $T$ parameter can be made to be  small if the custodial $SU(2)$ is not broken by the vector-like fermions. In the limit of $k_U=k_D (k_N=k_E)$ and $h_U=h_D (h_N=h_E)$, indeed $\Delta T$ vanishes.
If $k_U \neq k_D$, the vector-like fermions positively contribute to $\Delta T$.

\begin{SCfigure}
\centering
\includegraphics[width=0.45\textwidth]{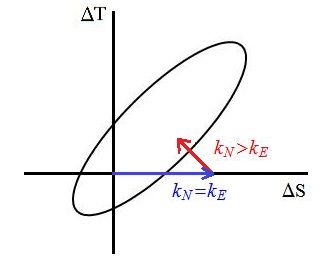}
\caption{Catoon picture of S and T from different choice of Yukawa couplings.
Although $k_N=k_E$ was taken for simplicity in the analysis of this paper,
it is possible to bring it back to 1 $\sigma$ ellipse of $\Delta S$ and $\Delta T$ by choosing 
a slightly large $k_N$ compared to $k_E$.
The choice of $k_N > k_E$ can generate positive $\Delta T$ and negative $\Delta S$. }
\label{ST}
\end{SCfigure}

The custodial $SU(2)$ limit ($k_N=k_E$) in our model is not the ideal point
for the electroweak precision observables
as $\Delta S$ monotonically grows as $M_F$ becomes light
while $\Delta T =0$ is maintained independently of $M_F$.
However, this problem can be easily overcome by making $k_E$ slightly larger than $k_N$.
The difference of $k_E$ and $k_N$ generates positive $\Delta T$.
Slightly large $k_E$ compared to $k_N$ will make the determinant of $E$ fermion to be smaller than that of $N$. As a result, the hypercharge ($Y=-\frac{1}{2}$) dependent term in $\Delta S$ becomes negative
and it can make $\Delta S$ to be smaller than the value computed in the $k_E=k_N$ limit.
Thus $k_E > k_N$ can provide nonzero positive $\Delta T$
and at the same time reduces $\Delta S$.

We presented the plots with the simplest choice which has the custodial $SU(2)$ symmetry.
As a result, the analytic expression became very simple. The general discussion is valid
for $k_E \neq k_N$ and to accommodate the electoweak precision observables is not difficult by allowing more general choices of parameters.

\section{Conclusions}

We considered a possibility of obtaining natural electroweak symmetry breaking
in the setup of large $\mu$ in the MSSM with vector-like matters.
Even for $\tan \beta =1$ at the scale $f \sim \mu \sim m_{\rm soft}$, 
the radiative correction of top quark and extra vector-like matters can explain
the observed 125 GeV mass of the Higgs boson.

Roughly, $m_h^2 \simeq 2 M_Z^2$ holds for $m_h = 125$ GeV.
In the MSSM with maximal stop mixing, tree level Higgs mass is $m_h^2 = M_Z^2$.
One loop correction from top/stop provides $\delta m_h^2 \sim M_Z^2$ if stop is at or below TeV
and the mixing is maximal.
In our setup, large $\mu$ and $\tan \beta \sim 1$ easily provide maximal stop mixing
for $\mu \sim 2$ TeV and $m_{\rm soft} \sim 1$ TeV.
For $\tan \beta = 1$, the tree level Higgs mass is zero.
Instead, extra vector-like fermions provide the additional $\delta m_h^2 \sim M_Z^2$
if the vector-like scalar partners are at around TeV.
As a result, $m_h^2 = (125 {\ \rm GeV})^2 \simeq 2 M_Z^2$ can be obtained.
One half is from top/stop loop and the other half is from vector-like matter loops.

The usual wisdom is that $\mu$ should be small to have a natural electroweak symmetry breaking.
Large $\mu$ can be compatible with the natural electroweak symmetry breaking
as long as the Higgs arises as a pseudo-Goldstone boson in supersymmetry
after the generation of large $\mu$ and $B\mu$.

We studied the Higgs phenomenology of $\tan \beta =1$ and large $\mu$ in the MSSM
with vector-like matters.
However, it would not be so difficult to find a UV completion of the model which naturally provides 
the massless scalar degree of freedom at a few TeV.
We leave the specific model construction of the Goldstone boson idea to future work.

The vector-like fermions with (large) order one Yukawa couplings make the Higgs quartic to be negative at (multi-)TeV scale. This feature was regarded as a dangerous aspect of models
explaining the enhanced di-photon rate. We showed that indeed it is a virtue
if the Higgs boson arises as a pseudo-Goldstone boson at (multi-)TeV scale.

We considered uncolored vector-like matters motivated by the enhanced Higgs to di-photon rate
of the LHC measurement.
It is possible to obtain 1.5 (1.2) times the Standard Model rate if the extra vector-like lepton is as light as 130 (250) GeV.
Indeed there is a correlation between the Higgs mass correction from the vector-like states
and the di-photon enhancement. If the soft supersymmetry breaking parameters are constrained to be smaller than 1 TeV (1.4 TeV), the Higgs to di-photon rate is predicted to be enhanced at least by factor about 1.6 (1.3) for $k=1$ and $N=1$. The minimum enhancement is obtained from the upper bound for $M_F$ needed to raise the Higgs mass for given $m_s$ assuming $\mu$ for the maximal mixing. As a result the minimum enhancement rate disappears for large $m_s$.

We leave the collider phenomenology of the vector-like states as a future work
which crucially depends on whether there is a mixed coupling of the vector-like states
with the Standard Model states.

\label{sec:conclusions}

\begin{acknowledgements}
This work is supported by the NRF of Korea No. 2011-0017051.
HD Kim thanks  CERN TH Institute, ``Beyond the Standard Model'' and KITPC program, ``The first two years of the LHC'' during which this work is completed.

\end{acknowledgements}

\end{document}